\documentclass[12pt]{article}
\pdfoutput=1

\usepackage{amsmath}
\usepackage{amssymb}
\usepackage{epsfig}
\usepackage{epstopdf}
\usepackage{latexsym}
\usepackage{color}
\numberwithin{equation}{section}
\usepackage[
      colorlinks=false,
      linkcolor=darkblue,  
      urlcolor=blue,    
      filecolor=blue,     
      citecolor=red,
linktocpage=true,
      pdfstartview=FitV,
      bookmarksopen=true    
      ]{hyperref}

\begin{document}

\begin{titlepage}

\centerline{\Huge \rm } 
\bigskip
\centerline{\Huge \rm Supersymmetric $AdS_6$ black holes}
\bigskip
\centerline{\Huge \rm from $F(4)$ gauged supergravity}
\bigskip
\bigskip
\bigskip
\bigskip
\bigskip
\centerline{\rm Minwoo Suh}
\bigskip
\centerline{\it Department of Physics, Kyungpook National University, Daegu 41566, Korea}
\bigskip
\centerline{\tt minwoosuh1@gmail.com} 
\bigskip
\bigskip
\bigskip
\bigskip
\bigskip
\bigskip
\bigskip
\bigskip
\bigskip
\bigskip
\bigskip
\bigskip

\begin{abstract}
\noindent In $F(4)$ gauged supergravity in six dimensions, we study supersymmetric $AdS_6$ black holes with various horizon geometries. We find a new $AdS_2\,\times\,\Sigma_{\mathfrak{g}_1}\times\Sigma_{\mathfrak{g}_2}$ horizon solution with $\mathfrak{g}_1>1$ and $\mathfrak{g}_2>1$, and present the black hole solution numerically. The full black hole is an interpolating geometry between the asymptotically $AdS_6$ boundary and the $AdS_2\,\times\,\Sigma_{\mathfrak{g}_1}\times\Sigma_{\mathfrak{g}_2}$ horizon. We calculate the Bekenstein-Hawking entropy of the black hole and find a match with the recently calculated topologically twisted index of 5d $USp(2N)$ gauge theory on $\Sigma_{\mathfrak{g}_1}\times\Sigma_{\mathfrak{g}_2}\times{S}^1$ in the large $N$ limit. We also find black hole horizons of K\"ahler four-cycles in Calabi-Yau fourfolds and on Cayley four-cycles in $Spin(7)$ manifolds.
\end{abstract}

\vskip 5cm

\flushleft {September, 2018}

\end{titlepage}

\tableofcontents

\section{Introduction}

The AdS$_6$/CFT$_5$ correspondence remains as one of the less appreciated among its family \cite{Maldacena:1997re}. In $SU(2)\times{U}(1)$-gauged $\mathcal{N}\,=\,4$ supergravity in six dimensions, commonly known as $F(4)$ gauged supergravity named after the $F(4)$ superalgebra in six dimensions, there is a unique supersymmetric fixed point \cite{Romans:1985tw}. This fixed point is known to be dual of 5d superconformal $USp(2N)$ gauge theory \cite{Ferrara:1998gv} which is one of the few 5d SCFTs known so far \cite{Seiberg:1996bd, Intriligator:1997pq}. In \cite{Cvetic:1999un} it was shown that $F(4)$ gauged supergravity is a consistent truncation of massive type IIA supergravity \cite{Romans:1985tz}.{\footnote {It is also a consistent truncation of type IIB supergravity \cite{Jeong:2013jfc, Hong:2018amk, Malek:2018zcz}.} The fixed point uplifts to $AdS_6\times{S}^4$ near-horizon geometry of the D4-D8 brane system \cite{Brandhuber:1999np}.

In order to study RG flows from 5d SCFTs to lower dimensional ones via the AdS/CFT correspondence in the spirit of \cite{Maldacena:2000mw}, twisted compactifications of $F(4)$ gauged supergravity were studied. The supergravity solutions describe the near-horizon geometries of wrapped D4-branes on various supersymmetric cycles. D4-branes wrapped on two- and three-cycles were studied in \cite{Nunez:2001pt, Naka:2002jz}. They found $AdS_4$ and $AdS_3$ fixed point solutions. See for more recent results on three-cycles in \cite{Dibitetto:2018iar}.

In this paper, we study supersymmetric $AdS_6$ black hole solutions by considering D4-branes wrapped on supersymmetric four-cycles. We begin by deriving supersymmetry equations for D4-branes wrapped on two Riemann surfaces and find a new $AdS_2\,\times\,H^2\,\times\,H^2$ solution. When we consider $S^2\,\times\,S^2$ background, there is no $AdS_2$ fixed point. We also present the full black hole solution numerically, which is an interpolating geometry between the asymptotically $AdS_6$ boundary and the $AdS_2\,\times\,H^2\,\times\,H^2$ horizon. We calculate entropy of asymptotically $AdS_6$ black holes with $AdS_2\,\times\,\Sigma_{\mathfrak{g}_1}\times\Sigma_{\mathfrak{g}_2}$ horizon where $\mathfrak{g}_1>1$ and $\mathfrak{g}_2>1$. Here $\Sigma_{\mathfrak{g}}$ is a Riemann surface of genus $\mathfrak{g}$ with constant curvature. Analogous to the 3d gauge theory examples in \cite{Cacciatori:2009iz, Benini:2015noa, Benini:2015eyy}, this entropy would give the topologically twisted index of 5d $USp(2N)$ gauge theory on $\Sigma_{\mathfrak{g}_1}\times\Sigma_{\mathfrak{g}_2}\times{S}^1$ in the large $N$ limit. Indeed we find that the entropy nicely matches the recent calculation of topologically twisted index in \cite{Hosseini:2018uzp}.

In contrast to the black hole solutions which exist only for $\mathfrak{g}_1>1$ and $\mathfrak{g}_2>1$, topologically twisted index is well defined for arbitrary genus. When a black hole solution exists, the index is counting the entropy of the black hole, however, the inverse is not true, $i.e.$ even if the index is well defined, it does not guarantee that there is a corresponding black hole solution. The index in this case might count some other information then entropy or might not be consistent.{\footnote {We would like to thank Alberto Zaffaroni for discussion on this point.}}

Then, we study supersymmetric $AdS_6$ black hole solutions with other horizon geometries by considering D4-branes wrapped on K\"ahler four-cycles in Calabi-Yau fourfolds and on Cayley four-cycles in $Spin(7)$ manifolds. We believe these are all possible four-cycles on which D4-branes can wrap. A product of two Riemann surfaces falls into a special case of K\"ahler four-cycles in Calabi-Yau fourfolds. We derive the supersymmetry equations, and obtain new supersymmetric $AdS_2$ solutions for each case.{\footnote {These cases were previously studied in \cite{Naka:2002jz} as D4-branes wrapped on K\"ahler four-cycles in Calabi-Yau threefolds and on co-associative four-cycles in $G_2$ manifolds. However, as they have not turned on the two form gauge potential which is needed to have a consistent set of supersymmetry equations and to satisfy the equations of motion, we conclude that their equations and solutions are $not$ correct.}}

We comment on the comparison with the recent field theory results. The topologically twisted index can be written as the contour integral of meromorphic differential form in variables parametrizing the Cartan subgroup and subalgebra of the gauge group, summed over the lattice of gauge magnetic fluxes, $\mathfrak{m}$, on the internal manifold \cite{Benini:2015noa, Benini:2015eyy}. Recently, topologically twisted index of 5d $USp(2N)$ gauge theory on $\Sigma_{\mathfrak{g}_1}\times\Sigma_{\mathfrak{g}_2}\times{S}^1$ was calculated at finite $N$ by two independent groups, \cite{Hosseini:2018uzp} and \cite{Crichigno:2018adf}, and their calculations agree. When considering the large $N$ limit, they both employed a conjecture to extremize the prepotential in order to get a saddle point distribution for dominant eigenvalue distribution. In \cite{Crichigno:2018adf}, by considering the non-zero magnetic fluxes as a subleading contribution, the contribution to the twisted index from zero magnetic fluxes, $\mathfrak{m}\,=\,0$, was evaluated. As we explain in detail below, it matches the gravitational entropy of the $AdS_2$ solution in \cite{Naka:2002jz} which is expected to be incorrect, and it counts the half of the full contribution to the twisted index. However, in \cite{Hosseini:2018uzp}, instead of setting the magnetic fluxes to zero, they extremized the twisted superpotential in order to get a saddle point distribution for the fluxes as well. In this paper, we will show that the full gravitational entropy of asymptotically $AdS_6$ black holes with $AdS_2\,\times\,\Sigma_{\mathfrak{g}_1}\times\Sigma_{\mathfrak{g}_2}$ horizon matches the topologically twisted index in the large $N$ limit, only when the contributions from non-zero gauge theory magnetic fluxes are accounted as it was done in \cite{Hosseini:2018uzp}. 

In section 2, we review $F(4)$ gauged supergravity. In section 3, we study supersymmetric black holes with $AdS_2\,\times\,\Sigma_{\mathfrak{g}_1}\times\Sigma_{\mathfrak{g}_2}$ horizon by considering D4-branes wrapped on two Riemann surfaces. We derive the supersymmetry equations, find a new $AdS_2$ solution, and calculate entropy of the black holes. We also present the full black hole solution numerically. In section 4, we consider more horizon geometries by studying D4-branes wrapped on K\"ahler four-cycles in Calabi-Yau fourfolds and on Cayley four-cycles in $Spin(7)$ manifolds. In section 5, we conclude. The equations of motion of $F(4)$ gauged supergravity are presented in appendix A.

\section{$F(4)$ gauged supergravity in six dimensions}

We review $SU(2)\times{U}(1)$-gauged $\mathcal{N}\,=\,4$ supergravity in six dimensions \cite{Romans:1985tw}. The bosonic field content consists of the metric, $g_{\mu\nu}$, a real scalar, $\phi$, an $SU(2)$ gauge field, $A^I_\mu$, $I\,=\,1,\,2,\,3$, a $U(1)$ gauge field, $\mathcal{A}_\mu$, and a two-form gauge potential, $B_{\mu\nu}$. The fermionic field content is gravitinos, $\psi_{\mu{i}}$, and dilatinos, $\chi_i$, $i\,=\,1,\,2$. The field strengths are defined by
\begin{align}
\mathcal{F}_{\mu\nu}\,=&\,\partial_\mu\mathcal{A}_\nu-\partial_\nu\mathcal{A}_\mu\,, \notag \\
F^I_{\mu\nu}\,=&\,\partial_\mu{A}^I_\nu-\partial_\nu{A}^I_\mu+g\epsilon^{IJK}A^J_\mu{A}^K_\nu\,, \notag \\
G_{\mu\nu\rho}\,=&\,3\partial_{[\mu}B_{\nu\rho]}\,, \notag \\
\mathcal{H}_{\mu\nu}\,=&\,\mathcal{F}_{\mu\nu}+mB_{\mu\nu}\,.
\end{align}
The bosonic Lagrangian is given by
\begin{align}
e^{-1}\mathcal{L}\,=\,&-\frac{1}{4}R+\frac{1}{2}\partial_\mu\phi\partial^\mu\phi+\frac{1}{8}\left(g^2e^{\sqrt{2}\phi}+4gme^{-\sqrt{2}\phi}-m^2e^{-3\sqrt{2}\phi}\right) \notag \\
&-\frac{1}{4}e^{-\sqrt{2}\phi}\left(\mathcal{H}_{\mu\nu}\mathcal{H}^{\mu\nu}+F^I_{\mu\nu}F^{I\mu\nu}\right)+\frac{1}{12}e^{2\sqrt{2}\phi}G_{\mu\nu\rho}G^{\mu\nu\rho} \notag \\
&-\frac{1}{8}\epsilon^{\mu\nu\rho\sigma\tau\kappa}B_{\mu\nu}\left(\mathcal{F}_{\rho\sigma}\mathcal{F}_{\tau\kappa}+mB_{\rho\sigma}\mathcal{F}_{\tau\kappa}+\frac{1}{3}m^2B_{\rho\sigma}B_{\tau\kappa}+F^I_{\rho\sigma}F^I_{\tau\kappa}\right)\,,
\end{align}
where $g$ is the $SU(2)$ gauge coupling constant and $m$ is the mass of the two-form gauge potential. The supersymmetry transformations of the fermionic fields are
\begin{align}
\delta\psi_{\mu{i}}\,=&\,\nabla_\mu\epsilon_i+gA^I_\mu(T^I)_i\,^j\epsilon_j-\frac{1}{8\sqrt{2}}\left(ge^{-\frac{\phi}{\sqrt{2}}}+me^{-\frac{3\phi}{\sqrt{2}}}\right)\gamma_\mu\gamma_7\epsilon_i \notag \\
&-\frac{1}{8\sqrt{2}}e^{-\frac{\phi}{\sqrt{2}}}\left(\mathcal{F}_{\nu\lambda}+mB_{\nu\lambda}\right)\left(\gamma_\mu\,^{\nu\lambda}-6\delta_\mu\,^\nu\gamma^\lambda\right)\epsilon_i \notag \\
&-\frac{1}{4\sqrt{2}}e^{-\frac{\phi}{\sqrt{2}}}F^I_{\nu\lambda}\left(\gamma_\mu\,^{\nu\lambda}-6\delta_\mu\,^\nu\gamma^\lambda\right)\gamma_7(T^I)_i\,^j\epsilon_j \notag \\
&-\frac{1}{24}e^{\sqrt{2}\phi}G_{\nu\lambda\rho}\gamma_7\gamma^{\nu\lambda\rho}\gamma_\mu\epsilon_i\,, \\
\delta\chi_i\,=&\,\frac{1}{\sqrt{2}}\gamma^\mu\partial_\mu\phi\epsilon_i+\frac{1}{4\sqrt{2}}\left(ge^{-\frac{\phi}{\sqrt{2}}}-3me^{-\frac{3\phi}{\sqrt{2}}}\right)\gamma_7\epsilon_i \notag \\
&+\frac{1}{4\sqrt{2}}e^{-\frac{\phi}{\sqrt{2}}}\left(\mathcal{F}_{\mu\nu}+mB_{\mu\nu}\right)\gamma^{\mu\nu}\epsilon_i \notag \\
&+\frac{1}{2\sqrt{2}}e^{-\frac{\phi}{\sqrt{2}}}F^I_{\mu\nu}\gamma^{\mu\nu}\gamma_7(T^I)_i\,^j\epsilon_j \notag \\
&-\frac{1}{12}e^{\sqrt{2}\phi}G_{\mu\nu\lambda}\gamma_7\gamma^{\mu\nu\lambda}\epsilon_i\,,
\end{align}
where $T^I$, $I$ = 1, 2, 3, are the $SU(2)$ left-invariant one-forms,
\begin{equation}
T^I\,=\,-\frac{i}{2}\sigma^I\,.
\end{equation}
Described by the above Lagrangian, there are five inequivalent theories : $\mathcal{N}\,=\,4^+$ ($g>0$, $m>0$), $\mathcal{N}\,=\,4^-$ ($g<0$, $m>0$), $\mathcal{N}\,=\,4^g$ ($g>0$, $m=0$), $\mathcal{N}\,=\,4^m$ ($g=0$, $m>0$), $\mathcal{N}\,=\,4^0$ ($g=0$, $m=0$). The $\mathcal{N}\,=\,4^+$ theory admits a supersymmetric $AdS_6$ fixed point when $g\,=\,3m$. At the supersymmetric $AdS_6$ fixed point, all the fields are vanishing except the $AdS_6$ metric.

\section{Black holes with $AdS_2\,\times\,\Sigma_{\mathfrak{g}_1}\times\Sigma_{\mathfrak{g}_2}$ horizon}

\subsection{The supersymmetry equations}

In this section, we obtain supersymmetric $AdS_6$ black holes with a horizon which is a product of two Riemann surfaces. We consider the metric,
\begin{equation}
ds^2\,=\,e^{2f(r)}\left(dt^2-dr^2\right)-e^{2g_1(r)}\left(d\theta_1^2+\sin^2\theta_1d\phi_1^2\right)-e^{2g_2(r)}\left(d\theta_2^2+\sin^2\theta_1d\phi_2^2\right)\,,
\end{equation}
for the $S^2\,\times\,S^2$ background, and
\begin{equation}
ds^2\,=\,e^{2f(r)}\left(dt^2-dr^2\right)-e^{2g_1(r)}\left(d\theta_1^2+\sinh^2\theta_1d\phi_1^2\right)-e^{2g_2(r)}\left(d\theta_2^2+\sinh^2\theta_1d\phi_2^2\right)\,,
\end{equation}
for the $H^2\,\times\,H^2$ background. The only non-vanishing component of the non-Abelian $SU(2)$ gauge field, $A^I_\mu$, $I$ = 1, 2, 3, is given by
\begin{equation}
A^3\,=\,-a_1\cos\theta_1{d}\phi_1-a_2\cos\theta_2{d}\phi_2\,,
\end{equation}
for the $S^2\,\times\,S^2$ background, and
\begin{equation}
A^3\,=\,a_1\cosh\theta_1{d}\phi_1+a_2\cosh\theta_2{d}\phi_2\,,
\end{equation}
for the $H^2\,\times\,H^2$ background, where the magnetic charges, $a_1$ and $a_2$, are constant. In order to have equal signs for field strengths, we set opposite signs of the gauge fields for $S^2\,\times\,S^2$ and $H^2\,\times\,H^2$ backgrounds. We also have a non-trivial two-form gauge potential, $B_{\mu\nu}$, and we will determine it later. We turn off the Abelian $U(1)$ gauge field, $\mathcal{A}_\mu$.

The supersymmetry equations are obtained by setting the supersymmetry variations of the fermionic fields to zero. From the supersymmetry variations, we obtain
\begin{align} \label{oneone}
f'e^{-f}\gamma^{\hat{r}}\epsilon_i&-\frac{1}{4\sqrt{2}}\left(ge^{\frac{\phi}{\sqrt{2}}}+me^{-\frac{3\phi}{\sqrt{2}}}\right)\gamma_7\epsilon_i \notag \\ -\frac{1}{2\sqrt{2}}e^{-\frac{\phi}{\sqrt{2}}}&\left(a_1e^{-2g_1}\gamma^{\hat{\theta_1}\hat{\phi_1}}+a_2e^{-2g_2}\gamma^{\hat{\theta_2}\hat{\phi_2}}\right)\gamma_72(T^3)_i\,^j\epsilon_j-\frac{3}{\sqrt{2}m}a_1a_2e^{\frac{\phi}{\sqrt{2}}-2g_1-2g_2}\gamma^{\hat{t}\hat{r}}\epsilon_i\,=\,0\,,\notag \\ 
\\ \label{twotwo}
g_1'e^{-f}\gamma^{\hat{r}}\epsilon_i&-\frac{1}{4\sqrt{2}}\left(ge^{\frac{\phi}{\sqrt{2}}}+me^{-\frac{3\phi}{\sqrt{2}}}\right)\gamma_7\epsilon_i \notag \\ +\frac{1}{2\sqrt{2}}e^{-\frac{\phi}{\sqrt{2}}}&\left(3a_1e^{-2g_1}\gamma^{\hat{\theta_1}\hat{\phi_1}}-a_2e^{-2g_2}\gamma^{\hat{\theta_2}\hat{\phi_2}}\right)\gamma_72(T^3)_i\,^j\epsilon_j+\frac{1}{\sqrt{2}m}a_1a_2e^{\frac{\phi}{\sqrt{2}}-2g_1-2g_2}\gamma^{\hat{t}\hat{r}}\epsilon_i\,=\,0\,, \\ \label{threethree}
g_2'e^{-f}\gamma^{\hat{r}}\epsilon_i&-\frac{1}{4\sqrt{2}}\left(ge^{\frac{\phi}{\sqrt{2}}}+me^{-\frac{3\phi}{\sqrt{2}}}\right)\gamma_7\epsilon_i \notag \\ +\frac{1}{2\sqrt{2}}e^{-\frac{\phi}{\sqrt{2}}}&\left(3a_2e^{-2g_2}\gamma^{\hat{\theta_2}\hat{\phi_2}}-a_1e^{-2g_1}\gamma^{\hat{\theta_1}\hat{\phi_1}}\right)\gamma_72(T^3)_i\,^j\epsilon_j+\frac{1}{\sqrt{2}m}a_1a_2e^{\frac{\phi}{\sqrt{2}}-2g_1-2g_2}\gamma^{\hat{t}\hat{r}}\epsilon_i\,=\,0\,, \\ \label{fourfour}
\frac{1}{\sqrt{2}}\phi'e^{-f}\gamma^{\hat{r}}\epsilon_i&+\frac{1}{4\sqrt{2}}\left(ge^{\frac{\phi}{\sqrt{2}}}-3me^{-\frac{3\phi}{\sqrt{2}}}\right)\gamma_7\epsilon_i \notag \\ +\frac{1}{2\sqrt{2}}e^{-\frac{\phi}{\sqrt{2}}}&\left(a_1e^{-2g_1}\gamma^{\hat{\theta_1}\hat{\phi_1}}+a_2e^{-2g_2}\gamma^{\hat{\theta_2}\hat{\phi_2}}\right)\gamma_72(T^3)_i\,^j\epsilon_j-\frac{1}{\sqrt{2}m}a_1a_2e^{\frac{\phi}{\sqrt{2}}-2g_1-2g_2}\gamma^{\hat{t}\hat{r}}\epsilon_i\,=\,0\,,
\end{align}
where the hatted indices are the flat indices. The $t$-, $\theta_1$-, and $\theta_2$-components of the gravitino variations give \eqref{oneone}, \eqref{twotwo}, \eqref{threethree}, and the dilatino variation gives \eqref{fourfour}. The $\phi_1$-, $\phi_2$-components of the variations are identical to the $\theta_1$-, and $\theta_2$-components beside few more terms,
\begin{equation} \label{pretwist}
\epsilon_i\,=\,-2ga_1\gamma^{\hat{\theta_1}\hat{\phi_1}}(T^3)_i\,^j\epsilon_j\,, \qquad \epsilon_i\,=\,-2ga_2\gamma^{\hat{\theta_2}\hat{\phi_2}}(T^3)_i\,^j\epsilon_j\,.
\end{equation}

We employ the projection conditions,
\begin{equation}
\gamma^{\hat{r}}\gamma^7\epsilon_i\,=\,\epsilon_i\,, \qquad \gamma^{\hat{\theta_1}\hat{\phi_1}}(T^3)_i\,^j\epsilon_j\,=\,\frac{\lambda}{2}\epsilon_i\,, \qquad \gamma^{\hat{\theta_2}\hat{\phi_2}}(T^3)_i\,^j\epsilon_j\,=\,\frac{\lambda}{2}\epsilon_i\,,
\end{equation}
where $\lambda\,=\,\pm{1}$. Solutions with the projection conditions preserve $1/8$ of the supersymmetries. By employing the projection conditions, we obtain the complete supersymmetry equations,
\begin{align} 
f'e^{-f}\,=&\,-\frac{1}{4\sqrt{2}}\left(ge^{\frac{\phi}{\sqrt{2}}}+me^{-\frac{3\phi}{\sqrt{2}}}\right)-\frac{\lambda}{2\sqrt{2}}e^{-\frac{\phi}{\sqrt{2}}}\left(a_1e^{-2g_1}+a_2e^{-2g_2}\right)-\frac{3}{\sqrt{2}m}a_1a_2e^{\frac{\phi}{\sqrt{2}}-2g_1-2g_2}\,, \notag \\ 
g_1'e^{-f}\,=&\,-\frac{1}{4\sqrt{2}}\left(ge^{\frac{\phi}{\sqrt{2}}}+me^{-\frac{3\phi}{\sqrt{2}}}\right)+\frac{\lambda}{2\sqrt{2}}e^{-\frac{\phi}{\sqrt{2}}}\left(3a_1e^{-2g_1}-a_2e^{-2g_2}\right)+\frac{1}{\sqrt{2}m}a_1a_2e^{\frac{\phi}{\sqrt{2}}-2g_1-2g_2}\,, \notag \\ 
g_2'e^{-f}\,=&\,-\frac{1}{4\sqrt{2}}\left(ge^{\frac{\phi}{\sqrt{2}}}+me^{-\frac{3\phi}{\sqrt{2}}}\right)+\frac{\lambda}{2\sqrt{2}}e^{-\frac{\phi}{\sqrt{2}}}\left(3a_2e^{-2g_2}-a_1e^{-2g_1}\right)+\frac{1}{\sqrt{2}m}a_1a_2e^{\frac{\phi}{\sqrt{2}}-2g_1-2g_2}\,, \notag \\ 
\frac{1}{\sqrt{2}}\phi'e^{-f}\,=&\,\frac{1}{4\sqrt{2}}\left(ge^{\frac{\phi}{\sqrt{2}}}-3me^{-\frac{3\phi}{\sqrt{2}}}\right)+\frac{\lambda}{2\sqrt{2}}e^{-\frac{\phi}{\sqrt{2}}}\left(a_1e^{-2g_1}+a_2e^{-2g_2}\right)-\frac{1}{\sqrt{2}m}a_1a_2e^{\frac{\phi}{\sqrt{2}}-2g_1-2g_2}\,.
\end{align}
From \eqref{pretwist} we also obtain the twist conditions on the magnetic charges,
\begin{equation} \label{twist}
a_1\,=\,-\frac{k}{\lambda{g}}\,, \qquad a_2\,=\,-\frac{k}{\lambda{g}}\,,
\end{equation}
where $k\,=\,+1$ for the $S^2\,\times\,S^2$ background and $k\,=\,-1$ for the $H^2\,\times\,H^2$ background. {\footnote {It is possible to have geometries like $S^2\times{H}^2$ for $k_1\,=\,+1$ and $k_2\,=\,-1$, or vice versa. One can easily generalize our supersymmetry equations and the twist conditions to that case.}}

In the derivation of the supersymmetry equations, we determined the non-zero components of the two-form gauge potential, $B_{\mu\nu}$. We determined the normalization by solving the equations of motion,
\begin{equation}
B_{tr}\,=\,-\frac{2}{m^2}a_1a_2e^{\sqrt{2}\phi+2f-2g_1-2g_2}\,.
\end{equation}
The three-form field strength of the two-form gauge potential, $G_{\mu\nu\lambda}$, vanishes identically. The supersymmetry equations satisfy the equations of motion. We present the equations of motion in appendix A.

When we plug the twist conditions, \eqref{twist}, in the supersymmetry equations, we obtain
\begin{align} \label{susy11}
f'e^{-f}\,=&\,-\frac{1}{4\sqrt{2}}\left(ge^{\frac{\phi}{\sqrt{2}}}+me^{-\frac{3\phi}{\sqrt{2}}}\right)+\frac{k}{2\sqrt{2}g}e^{-\frac{\phi}{\sqrt{2}}}\left(e^{-2g_1}+e^{-2g_2}\right)-\frac{3}{\sqrt{2}g^2m}e^{\frac{\phi}{\sqrt{2}}-2g_1-2g_2}\,, \notag \\ 
g_1'e^{-f}\,=&\,-\frac{1}{4\sqrt{2}}\left(ge^{\frac{\phi}{\sqrt{2}}}+me^{-\frac{3\phi}{\sqrt{2}}}\right)-\frac{k}{2\sqrt{2}g}e^{-\frac{\phi}{\sqrt{2}}}\left(3e^{-2g_1}-e^{-2g_2}\right)+\frac{1}{\sqrt{2}g^2m}e^{\frac{\phi}{\sqrt{2}}-2g_1-2g_2}\,, \notag \\ 
g_2'e^{-f}\,=&\,-\frac{1}{4\sqrt{2}}\left(ge^{\frac{\phi}{\sqrt{2}}}+me^{-\frac{3\phi}{\sqrt{2}}}\right)-\frac{k}{2\sqrt{2}g}e^{-\frac{\phi}{\sqrt{2}}}\left(3e^{-2g_2}-e^{-2g_1}\right)+\frac{1}{\sqrt{2}g^2m}e^{\frac{\phi}{\sqrt{2}}-2g_1-2g_2}\,, \notag \\ 
\frac{1}{\sqrt{2}}\phi'e^{-f}\,=&\,\frac{1}{4\sqrt{2}}\left(ge^{\frac{\phi}{\sqrt{2}}}-3me^{-\frac{3\phi}{\sqrt{2}}}\right)-\frac{k}{2\sqrt{2}g}e^{-\frac{\phi}{\sqrt{2}}}\left(e^{-2g_1}+e^{-2g_2}\right)-\frac{1}{\sqrt{2}g^2m}e^{\frac{\phi}{\sqrt{2}}-2g_1-2g_2}\,,
\end{align}
where $k\,=\,+1$ for the $S^2\,\times\,S^2$ background and $k\,=\,-1$ for the $H^2\,\times\,H^2$ background. The supersymmetry equations in \eqref{susy11} are analogous to the equations for M5-branes wrapped on two Riemann surfaces in \cite{Gauntlett:2001jj}, and more recelty generalized in \cite{Benini:2013cda}.

\subsection{The $AdS_2$ solution and entropy of black holes}

Now  we will consider the $\mathcal{N}\,=\,4^+$ theory, $g>0$, $m>0$. We find a new $AdS_2$ fixed point solution for the $H^2\,\times\,H^2$ background with $k\,=\,-1$,{\footnote {The solution can also be presented as
\begin{equation}
e^f\,=\,\frac{\sqrt{2}}{g}e^{-\frac{\phi}{\sqrt{2}}}\frac{1}{r}\,, \qquad e^{g_1}\,=\,e^{g_2}\,=\,\frac{2}{g}e^{-\frac{\phi}{\sqrt{2}}}\,, \qquad e^{-2\sqrt{2}\phi}\,=\,\frac{g}{2m}\,.
\end{equation}}}
\begin{equation} \label{ads2sol}
e^f\,=\,\frac{2^{1/4}}{g^{3/4}m^{1/4}}\frac{1}{r}\,, \qquad e^{g_1}\,=\,e^{g_2}\,=\,\frac{2^{3/4}}{g^{3/4}m^{1/4}}\,, \qquad e^{\frac{\phi}{\sqrt{2}}}\,=\,\frac{2^{1/4}m^{1/4}}{g^{1/4}}\,.
\end{equation}
When we consider the $S^2\,\times\,S^2$ background with $k\,=\,+1$, $AdS_2$ fixed point does not exist.

By employing the uplift formulae in \cite{Cvetic:1999un}, the $AdS_2$ solution can be uplifted to a solution of massive type IIA supergravity. We only present the uplift formulae for the metric,
\begin{equation}
ds_{10}^2\,=\,\sin^{1/12}\xi{X}^{1/8}\left[\Delta^{3/8}ds_6^2+\frac{2\Delta^{3/8}X^2}{g^2}d\xi^2+\frac{\cos^2\xi}{2g^2\Delta^{5/8}X}\Sigma^3_{I=1}\left(\sigma^I-gA^I\right)^2\right]\,,
\end{equation}
and the dilaton field, $\Phi$,
\begin{equation}
e^\Phi\,=\,\frac{\Delta^{1/4}}{\sin^{5/6}\xi{X}^{5/4}}\,,
\end{equation}
where we define
\begin{equation}
\Delta\,=\,X\cos^2\xi+\frac{1}{X^3}\sin^2\xi\,, \qquad X\,=\,e^{-\frac{\phi}{\sqrt{2}}}\,.
\end{equation}

The solution in \eqref{ads2sol} describes the $AdS_2\,\times\,H^2\times\,H^2$ horizon of six-dimensional black holes. Moreover, when we consider the $\mathcal{N}\,=\,4^+$ theory with $g\,=\,3m$, there is a supersymmetric $AdS_6$ fixed point which is known to be dual of 5d $USp(2N)$ gauge theory in the large $N$ limit. Recently, there has been development in calculating topologically twisted index of 5d $USp(2N)$ gauge theory on $M_4\times{S}^1$ by two independent groups, \cite{Hosseini:2018uzp} and \cite{Crichigno:2018adf}. See the introduction for more details on their calculations. Analogous to the 3d gauge theory examples in \cite{Cacciatori:2009iz, Benini:2015noa, Benini:2015eyy}, the entropy of the black holes could match the topologically twisted index of 5d $USp(2N)$ gauge theory on $\Sigma_{\mathfrak{g}_1}\times\Sigma_{\mathfrak{g}_2}\times{S}^1$ in the large $N$ limit. 

Now we calculate entropy of asymptotically $AdS_6$ black holes with $AdS_2\,\times\,H^2\times\,H^2$ horizon. It can be easily generalized to $AdS_2\,\times\,\Sigma_{\mathfrak{g}_1}\times\Sigma_{\mathfrak{g}_2}$ horizon where $\mathfrak{g}_1>1$ and $\mathfrak{g}_2>1$. Here $\Sigma_{\mathfrak{g}}$ is a Riemann surface of genus $\mathfrak{g}$ with constant curvature. We would like to consider the $AdS_6$ fixed point in the $\mathcal{N}\,=\,4^+$ theory by taking $g\,=\,3m$. In order to have unit radius, $L_{AdS_6}\,=\,1$, we set $m\,=\,\sqrt{2}$.{\footnote {See discussions around (4.6) in \cite{Bobev:2017uzs}.}}  The Bekenstein-Hawking entropy of the black hole is given by
\begin{equation}
S_{BH}\,=\,\frac{L_{AdS_p}^{p-2}}{4G_N^{(p)}}\,=\,\frac{1}{4G_N^{(2)}}\,=\,\frac{vol(M_4)}{4G_N^{(6)}}\,=\,\frac{e^{2g_1}vol{(\Sigma_{\mathfrak{g}_1})}e^{2g_2}vol{(\Sigma_{\mathfrak{g}_2})}}{4G_N^{(6)}}\,\,=\,\frac{8\pi^2(\mathfrak{g}_1-1)(\mathfrak{g}_2-1)}{27G_N^{(6)}}\,,
\end{equation}
where $L$, $vol$, and $G_N$ are radius, volume, and the Newton's gravitational constant for the corresponding spaces, respectively. We used that the volume of Riemann surfaces with genus, $\mathfrak{g}>1$, is{\footnote {See, for example, (73) in \cite{Maldacena:2000mw}.}}
\begin{equation}
vol(\Sigma_{\mathfrak{g}})\,=\,4\pi(\mathfrak{g}-1)\,,
\end{equation}
and, in the last equality, we used the value of the warp factor at our $AdS_2$ solution, $e^{4g}\,=\,2/3^3$. We can relate the gravitational entropy to the free energy of 5d SCFTs on $S^5$ by using the universal formula,{\footnote {See, for example, (4.5) in \cite{Bobev:2017uzs}.}}
\begin{equation}
F_{S^5}\,=\,-\frac{\pi^2L^4_{AdS_6}}{3G_N^{(6)}}\,,
\end{equation}
and then entropy of the black hole is
\begin{equation}
S_{BH}\,=\,-\frac{8}{9}(\mathfrak{g}_1-1)(\mathfrak{g}_2-1)F_{S^5}\,.
\end{equation}
The free energy of 5d $USp(2N)$ gauge theory with $N_f$ flavors on $S^5$ in the large $N$ limit is given by \cite{Jafferis:2012iv}, 
\begin{equation}
F_{S^5}\,=\,-\frac{9\sqrt{2}\pi{N}^{5/2}}{5\sqrt{8-N_f}}\,.
\end{equation}
Therefore, entropy of the black hole can be written by
\begin{equation}
S_{BH}\,=\,\frac{8\sqrt{2}\pi(\mathfrak{g}_1-1)(\mathfrak{g}_2-1)N^{5/2}}{5\sqrt{8-N_f}}\,,
\end{equation}
where $\mathfrak{g}_1>1$ and $\mathfrak{g}_2>1$. This nicely matches the topologically twisted index of 5d $USp(2N)$ gauge theory on $\Sigma_{\mathfrak{g}_1}\times\Sigma_{\mathfrak{g}_2}\times{S}^1$ in the large $N$ limit calculated in \cite{Hosseini:2018uzp}. As it was explained in the introduction, the contribution to the topologically twisted index with zero magnetic fluxes, $\mathfrak{m}\,=\,0$, calculated in \cite{Crichigno:2018adf}, counts only the half of the gravitational entropy.

\subsection{Numerical black hole solutions}

\begin{figure}[h!]
\begin{center}
\includegraphics[width=2.0in]{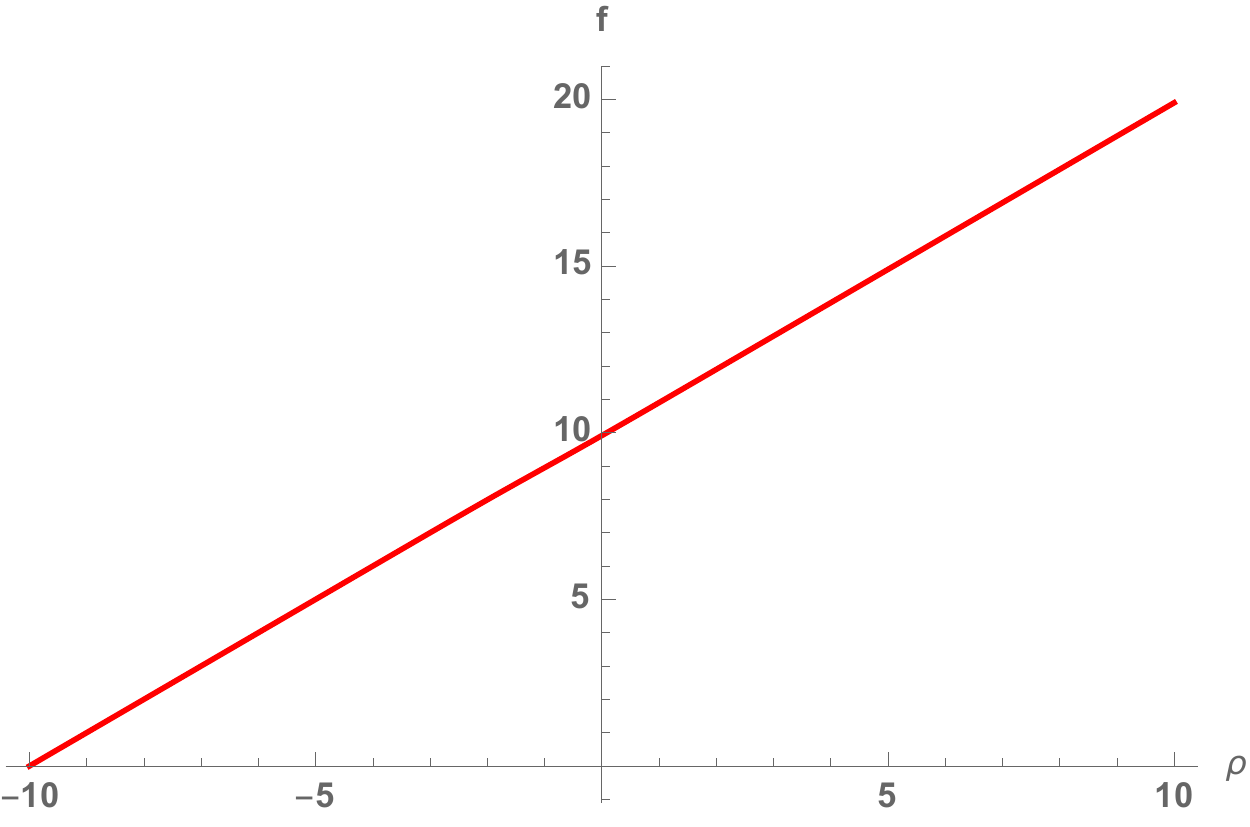} \qquad \includegraphics[width=2.0in]{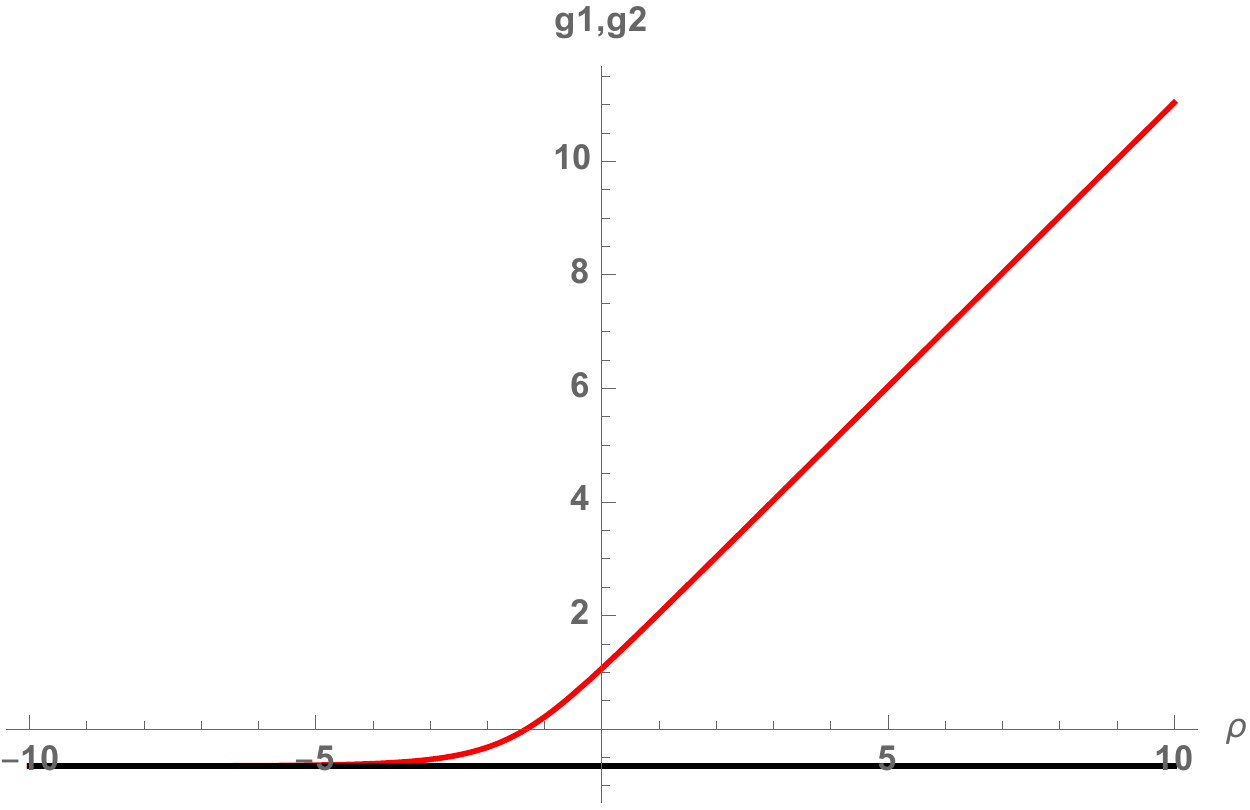} \qquad \includegraphics[width=2.0in]{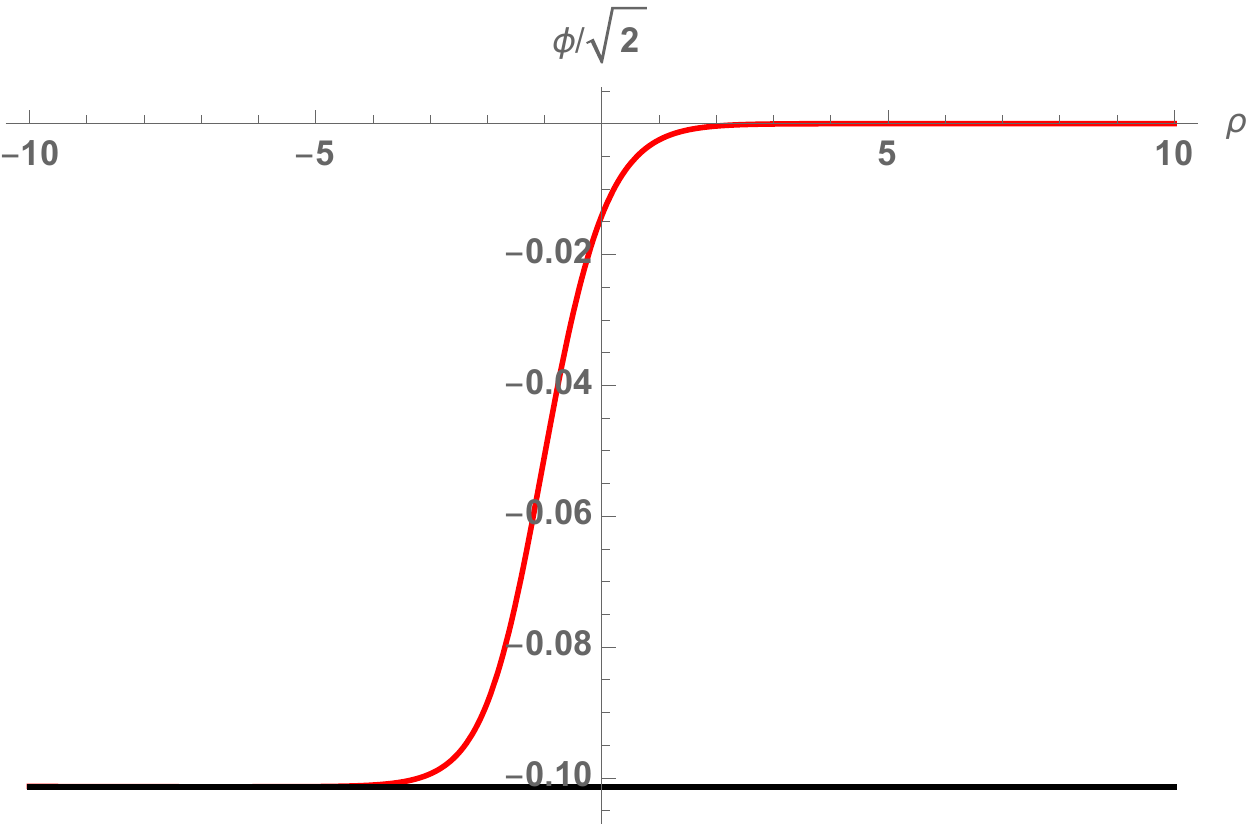}
\caption{{\it Numerical black hole solution with $m\,=\,\sqrt{2}$ and $g\,=\,3m$. The black straight lines are the values at the $AdS_2$ horizon given in \eqref{ads2sol}.}}
\label{1}
\end{center}
\end{figure}
Now we present the full black hole solution numerically. The full black hole solution is an interpolating geometry between the asymptotically $AdS_6$ boundary and the $AdS_2\,\times\,\Sigma_{\mathfrak{g}_1}\times\Sigma_{\mathfrak{g}_2}$ horizon. We introduce a new radial coordinate,
\begin{equation} \label{rhocoord}
\rho\,=\,f+\frac{\phi}{\sqrt{2}}\,.
\end{equation}
This kind of coordinate was introduced in \cite{Benini:2013cda}. Employing the supersymmetry equations, we obtain
\begin{equation}
\frac{\partial\rho}{\partial{r}}\,=\,f'+\frac{\phi'}{\sqrt{2}}\,=\,-e^fD\,,
\end{equation}
where we define
\begin{equation}
D\,=\,\frac{m}{\sqrt{2}}e^{-\frac{3\phi}{\sqrt{2}}}+\frac{4}{\sqrt{2}m}a_1a_2e^{\frac{\phi}{\sqrt{2}}-2g_1-2g_2}\,.
\end{equation}
Then, the supersymmetry equations are
\begin{align} 
-D\frac{\partial{f}}{\partial\rho}\,=&\,-\frac{1}{4\sqrt{2}}\left(ge^{\frac{\phi}{\sqrt{2}}}+me^{-\frac{3\phi}{\sqrt{2}}}\right)-\frac{\lambda}{2\sqrt{2}}e^{-\frac{\phi}{\sqrt{2}}}\left(a_1e^{-2g_1}+a_2e^{-2g_2}\right)-\frac{3}{\sqrt{2}m}a_1a_2e^{\frac{\phi}{\sqrt{2}}-2g_1-2g_2}\,, \notag \\ 
-D\frac{\partial{g_1}}{\partial\rho}\,=&\,-\frac{1}{4\sqrt{2}}\left(ge^{\frac{\phi}{\sqrt{2}}}+me^{-\frac{3\phi}{\sqrt{2}}}\right)+\frac{\lambda}{2\sqrt{2}}e^{-\frac{\phi}{\sqrt{2}}}\left(3a_1e^{-2g_1}-a_2e^{-2g_2}\right)+\frac{1}{\sqrt{2}m}a_1a_2e^{\frac{\phi}{\sqrt{2}}-2g_1-2g_2}\,, \notag \\ 
-D\frac{\partial{g_2}}{\partial\rho}\,=&\,-\frac{1}{4\sqrt{2}}\left(ge^{\frac{\phi}{\sqrt{2}}}+me^{-\frac{3\phi}{\sqrt{2}}}\right)+\frac{\lambda}{2\sqrt{2}}e^{-\frac{\phi}{\sqrt{2}}}\left(3a_2e^{-2g_2}-a_1e^{-2g_1}\right)+\frac{1}{\sqrt{2}m}a_1a_2e^{\frac{\phi}{\sqrt{2}}-2g_1-2g_2}\,, \notag \\ 
-\frac{1}{\sqrt{2}}D\frac{\partial\phi}{\partial\rho}\,=&\,\frac{1}{4\sqrt{2}}\left(ge^{\frac{\phi}{\sqrt{2}}}-3me^{-\frac{3\phi}{\sqrt{2}}}\right)+\frac{\lambda}{2\sqrt{2}}e^{-\frac{\phi}{\sqrt{2}}}\left(a_1e^{-2g_1}+a_2e^{-2g_2}\right)-\frac{1}{\sqrt{2}m}a_1a_2e^{\frac{\phi}{\sqrt{2}}-2g_1-2g_2}\,.
\end{align}
In the $r$-coordinate, the UV or asymptotically $AdS_6$ boundary is at $r\,=\,0$, and the IR or $AdS_2\,\times\,\Sigma_{\mathfrak{g}_1}\times\Sigma_{\mathfrak{g}_2}$ horizon is at $r\,=\,\infty$. In this $\rho$-coordinate, the UV is at $\rho\,=\,+\infty$, and the IR is at $\rho\,=\,-\infty$. We present the plot of the full black hole solution in figure 1. We have set $m\,=\,\sqrt{2}$ and $g\,=\,3m$, and there is no free parameter left. Therefore, there is only one plot.

\section{Black holes with other horizons}

In this section, we obtain more black hole solutions with other horizon geometries by considering D4-branes wrapped on K\"ahler four-cycles in Calabi-Yau fourfolds and on Cayley four-cycles in $Spin(7)$ manifolds. We believe these are all possible four-cycles on which D4-branes can wrap in $F(4)$ gauged supergravity. D4-branes on two Riemann surfaces in the previous section fall into a special case of D4-branes on K\"ahler four-cycles in Calabi-Yau fourfolds. The analogous results of M5-branes wrapped on supersymmetric four-cycles were studied in \cite{Gauntlett:2000ng, Gauntlett:2001jj}.

\subsection{K\"ahler four-cycles in Calabi-Yau fourfolds}

We consider the metric,
\begin{equation}
ds^2\,=\,e^{2f(r)}\left(dt^2-dr^2\right)-e^{2g(r)}ds^2_{M_4}\,,
\end{equation}
where $M_4$ is a K\"ahler four-cycle in Calabi-Yau fourfolds. The curved coordinates on the K\"ahler four-cycles will be denoted by $\{x_1,\,x_2,\,x_3,\,x_4\}$, and the hatted ones are the flat coordinates. For K\"ahler four-cycles in Calabi-Yau fourfolds, there are four directions transverse to D4-branes in the fourfolds. The normal bundle of the four-cycle has $U(2)\,\subset\,SO(4)$ structure group. We identify $U(1)$ part of the structure group with $U(1)$ gauge field from the non-Abelian $SU(2)$ gauge group, \cite{Gauntlett:2000ng, Gauntlett:1998vk}. The only non-vanishing component of the non-Abelian $SU(2)$ gauge field, $A^I_\mu$, $I$ = 1, 2, 3, is given by
\begin{equation}
F^3_{\hat{x}_1\hat{x}_2}\,=\,a_1e^{-2g}\,, \qquad F^3_{\hat{x}_3\hat{x}_4}\,=\,a_2e^{-2g}\,,
\end{equation}
where the magnetic charges, $a_1$ and $a_2$, are constant. The only non-vanishing component of the two-form gauge potential is
\begin{equation}
B_{tr}\,=\,-\frac{2}{m^2}a_1a_2e^{\sqrt{2}\phi+2f-4g}\,.
\end{equation}
We employ the projection conditions,
\begin{equation}
\gamma^{\hat{r}}\gamma^7\epsilon_i\,=\,\epsilon_i\,, \qquad \gamma^{\hat{x}_1\hat{x}_2}(T^3)_i\,^j\epsilon_j\,=\,\frac{\lambda}{2}\epsilon_i\,, \qquad \gamma^{\hat{x}_3\hat{x}_4}(T^3)_i\,^j\epsilon_j\,=\,\frac{\lambda}{2}\epsilon_i\,,
\end{equation}
where $\lambda\,=\,\pm{1}$. Solutions with the projection conditions preserve $1/8$ of the supersymmetries. By employing the projection conditions, we obtain the complete supersymmetry equations,
\begin{align} \label{susyk}
f'e^{-f}\,=&\,-\frac{1}{4\sqrt{2}}\left(ge^{\frac{\phi}{\sqrt{2}}}+me^{-\frac{3\phi}{\sqrt{2}}}\right)+\frac{k}{\sqrt{2}g}e^{-\frac{\phi}{\sqrt{2}}-2g}-\frac{3}{\sqrt{2}g^2m}e^{\frac{\phi}{\sqrt{2}}-4g}\,, \notag \\ 
g'e^{-f}\,=&\,-\frac{1}{4\sqrt{2}}\left(ge^{\frac{\phi}{\sqrt{2}}}+me^{-\frac{3\phi}{\sqrt{2}}}\right)-\frac{k}{\sqrt{2}g}e^{-\frac{\phi}{\sqrt{2}}-2g}+\frac{1}{\sqrt{2}g^2m}e^{\frac{\phi}{\sqrt{2}}-4g}\,, \notag \\ 
\frac{1}{\sqrt{2}}\phi'e^{-f}\,=&\,\frac{1}{4\sqrt{2}}\left(ge^{\frac{\phi}{\sqrt{2}}}-3me^{-\frac{3\phi}{\sqrt{2}}}\right)-\frac{k}{\sqrt{2}g}e^{-\frac{\phi}{\sqrt{2}}-2g}-\frac{1}{\sqrt{2}g^2m}e^{\frac{\phi}{\sqrt{2}}-4g}\,,
\end{align}
with the twist conditions,
\begin{equation}
a_1\,=\,-\frac{k}{\lambda{g}}\,, \qquad a_2\,=\,-\frac{k}{\lambda{g}}\,,
\end{equation}
where $k$ determines the curvature of the K\"ahler four-cycles in Calabi-Yau fourfolds. 

A product of two Riemann surfaces considered in the previous section is a special case of K\"ahler four-cycles in Calabi-Yau fourfolds. When we identify $g\,\equiv\,g_1\,=\,g_2$ in the supersymmetry equations for D4-branes wrapped on two Riemann surfaces, \eqref{susy11}, we obtain the supersymmetry equations here, \eqref{susyk}. By solving the supersymmetry equations, we find an $AdS_2$ fixed point solution which is identical to the one obtained in the previous section, \eqref{ads2sol}.

\subsection{Cayley four-cycles in $Spin(7)$ manifolds}

We consider the metric,
\begin{equation}
ds^2\,=\,e^{2f(r)}\left(dt^2-dr^2\right)-e^{2g(r)}ds^2_{M_4}\,,
\end{equation}
where $M_4$ is a Cayley four-cycle in manifolds with $Spin(7)$ holonomy. The curved coordinates on the Cayley four-cycles will be denoted by $\{x_1,\,x_2,\,x_3,\,x_4\}$, and the hatted ones are the flat coordinates. In order to preserve supersymmetry for D4-branes wrapped on Cayley four-cycles in $Spin(7)$ manifolds, we identify self-dual $SU(2)_+$ subgroup of the $SO(4)$ isometry of the four-cycle,
\begin{equation}
SO(4)\,\rightarrow\,SU(2)_+\,\times\,SU(2)_-\,,
\end{equation}
with the non-Abelian $SU(2)$ gauge group, \cite{Gauntlett:2000ng, Gauntlett:1998vk}. The self-duality is defined by
\begin{equation}
\gamma_{\mu\nu}\,=\,\pm\frac{1}{2}\epsilon_{\mu\nu\rho\sigma}\gamma^{\rho\sigma}\,,
\end{equation}
and we denoted the self-duality and anti-self-duality by $+$ and $-$, respectively. For the self-dual part, components are identified by
\begin{equation}
\gamma^{\hat{x}_1\hat{x}_2}\,=\,\gamma^{\hat{x}_3\hat{x}_4}\,, \qquad \gamma^{\hat{x}_1\hat{x}_3}\,=\,\gamma^{\hat{x}_4\hat{x}_2}\,, \qquad \gamma^{\hat{x}_1\hat{x}_4}\,=\,\gamma^{\hat{x}_2\hat{x}_3}\,.
\end{equation}
The only non-vanishing components of the non-Abelian $SU(2)$ gauge field, $A^I_\mu$, $I$ = 1, 2, 3, are given by
\begin{align}
F^1_{\hat{x}_1\hat{x}_2}\,=\,\,F^1_{\hat{x}_3\hat{x}_4}\,=\,a_1e^{-2g}\,, \notag \\
F^2_{\hat{x}_1\hat{x}_3}\,=\,\,F^2_{\hat{x}_4\hat{x}_2}\,=\,a_2e^{-2g}\,, \notag \\
F^3_{\hat{x}_1\hat{x}_4}\,=\,\,F^3_{\hat{x}_2\hat{x}_3}\,=\,a_3e^{-2g}\,,
\end{align}
where the magnetic charges, $a_1$, $a_2$ and $a_3$, are constant. The only non-vanishing component of the two-form gauge potential is
\begin{equation}
B_{tr}\,=\,-\frac{2}{m^2}\left(a_1^2+a_2^2+a_3^2\right)e^{\sqrt{2}\phi+2f-4g}\,.
\end{equation}
We employ the projection conditions,
\begin{align}
\gamma^{\hat{r}}\gamma^7&\epsilon_i\,=\,\epsilon_i\,, \notag \\
\gamma^{\hat{x}_1\hat{x}_2}(T^1)_i\,^j\epsilon_j\,=&\,\gamma^{\hat{x}_3\hat{x}_4}(T^1)_i\,^j\epsilon_j\,=\,\frac{\lambda}{2}\epsilon_i\,, \notag \\
\gamma^{\hat{x}_1\hat{x}_3}(T^2)_i\,^j\epsilon_j\,=&\,\gamma^{\hat{x}_4\hat{x}_2}(T^2)_i\,^j\epsilon_j\,=\,\frac{\lambda}{2}\epsilon_i\,, \notag \\
\gamma^{\hat{x}_1\hat{x}_4}(T^3)_i\,^j\epsilon_j\,=&\,\gamma^{\hat{x}_2\hat{x}_3}(T^3)_i\,^j\epsilon_j\,=\,\frac{\lambda}{2}\epsilon_i\,,
\end{align}
where $\lambda\,=\,\pm{1}$. Solutions with the projection conditions preserve $1/16$ of the supersymmetries. By employing the projection conditions, we obtain the complete supersymmetry equations,
\begin{align}
f'e^{-f}\,=&\,-\frac{1}{4\sqrt{2}}\left(ge^{\frac{\phi}{\sqrt{2}}}+me^{-\frac{3\phi}{\sqrt{2}}}\right)+\frac{k}{\sqrt{2}g}e^{-\frac{\phi}{\sqrt{2}}-2g}-\frac{1}{\sqrt{2}g^2m}e^{\frac{\phi}{\sqrt{2}}-4g}\,, \notag \\ 
g'e^{-f}\,=&\,-\frac{1}{4\sqrt{2}}\left(ge^{\frac{\phi}{\sqrt{2}}}+me^{-\frac{3\phi}{\sqrt{2}}}\right)-\frac{k}{\sqrt{2}g}e^{-\frac{\phi}{\sqrt{2}}-2g}+\frac{1}{3\sqrt{2}g^2m}e^{\frac{\phi}{\sqrt{2}}-4g}\,, \notag \\ 
\frac{1}{\sqrt{2}}\phi'e^{-f}\,=&\,\frac{1}{4\sqrt{2}}\left(ge^{\frac{\phi}{\sqrt{2}}}-3me^{-\frac{3\phi}{\sqrt{2}}}\right)-\frac{k}{\sqrt{2}g}e^{-\frac{\phi}{\sqrt{2}}-2g}-\frac{1}{3\sqrt{2}g^2m}e^{\frac{\phi}{\sqrt{2}}-4g}\,,
\end{align}
with the twist conditions,
\begin{equation}
a_1\,=\,-\frac{k}{3\lambda{g}}\,, \qquad a_2\,=\,-\frac{k}{3\lambda{g}}\,, \qquad a_3\,=\,-\frac{k}{3\lambda{g}}\,,
\end{equation}
where $k$ determines the curvature of the Cayley four-cycles in $Spin(7)$ manifolds.

Now  we will consider the $\mathcal{N}\,=\,4^+$ theory, $g>0$, $m>0$. By solving the supersymmetry equations, we find a new $AdS_2$ fixed point solution for the negatively curved Cayley four-cycles with $k\,=\,-1$,
\begin{equation} \label{cayleyads2}
e^f\,=\,\frac{3^{1/4}}{g^{3/4}m^{1/4}}\frac{1}{r}\,, \qquad e^{g}\,=\,\frac{2}{3^{1/4}g^{3/4}m^{1/4}}\,, \qquad e^{\frac{\phi}{\sqrt{2}}}\,=\,\frac{2^{1/2}m^{1/4}}{3^{1/4}g^{1/4}}\,.
\end{equation}
When we consider $k\,=\,+1$, $AdS_2$ fixed point does not exist. It will be interesting to have a  field theory interpretation of this $AdS_2$ fixed point solution.

Now we present the full black hole solution numerically. The full black hole solution is an interpolating geometry between the asymptotically $AdS_6$ boundary and the $AdS_2\,\times\,Cayley_4$ horizon with negative curvature, $k\,=\,-1$. We introduce the radial coordinate, $\rho$, in \eqref{rhocoord}, and solve the supersymmetry equations in $\rho$ coordinate, as we did in section 3.3. We again set $m\,=\,\sqrt{2}$ and $g\,=\,3m$. The full black hole solution is presented in figure 2.
\begin{figure}[h!]
\begin{center}
\includegraphics[width=2.0in]{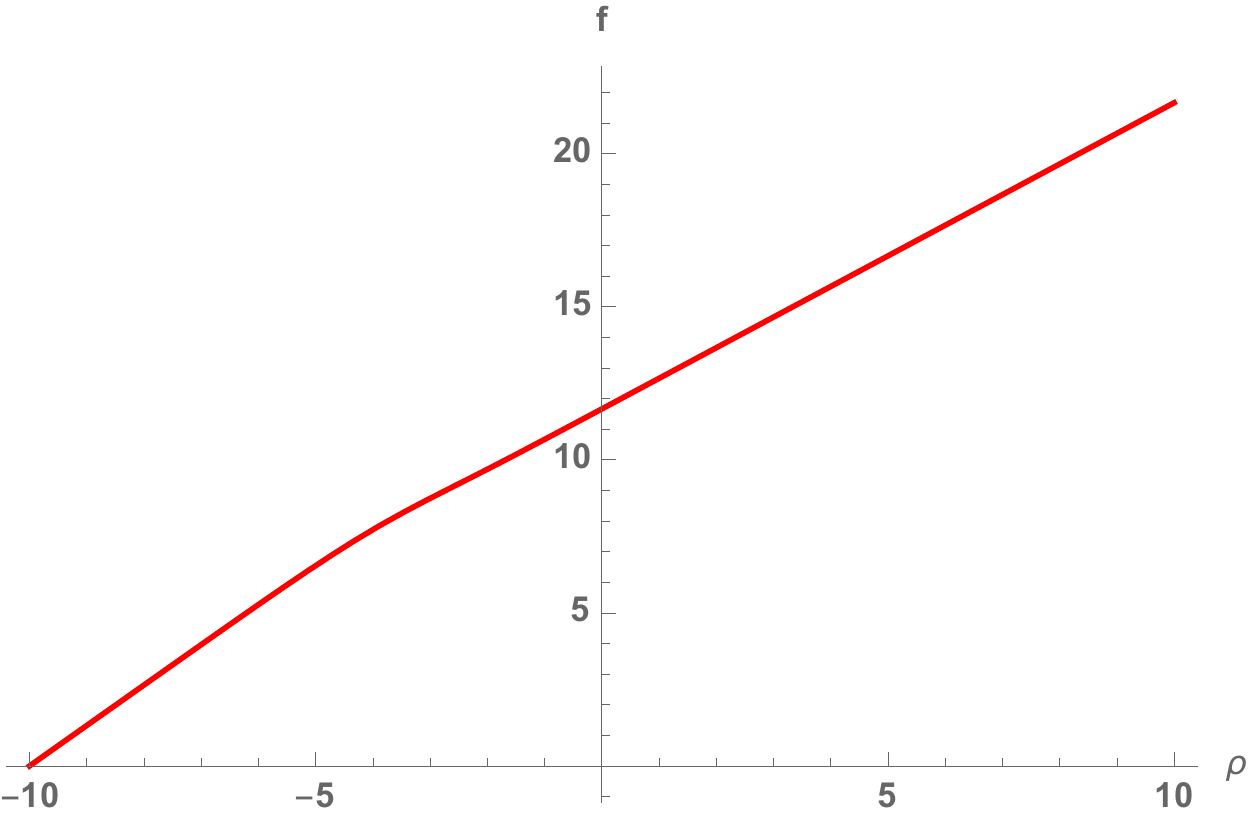} \qquad \includegraphics[width=2.0in]{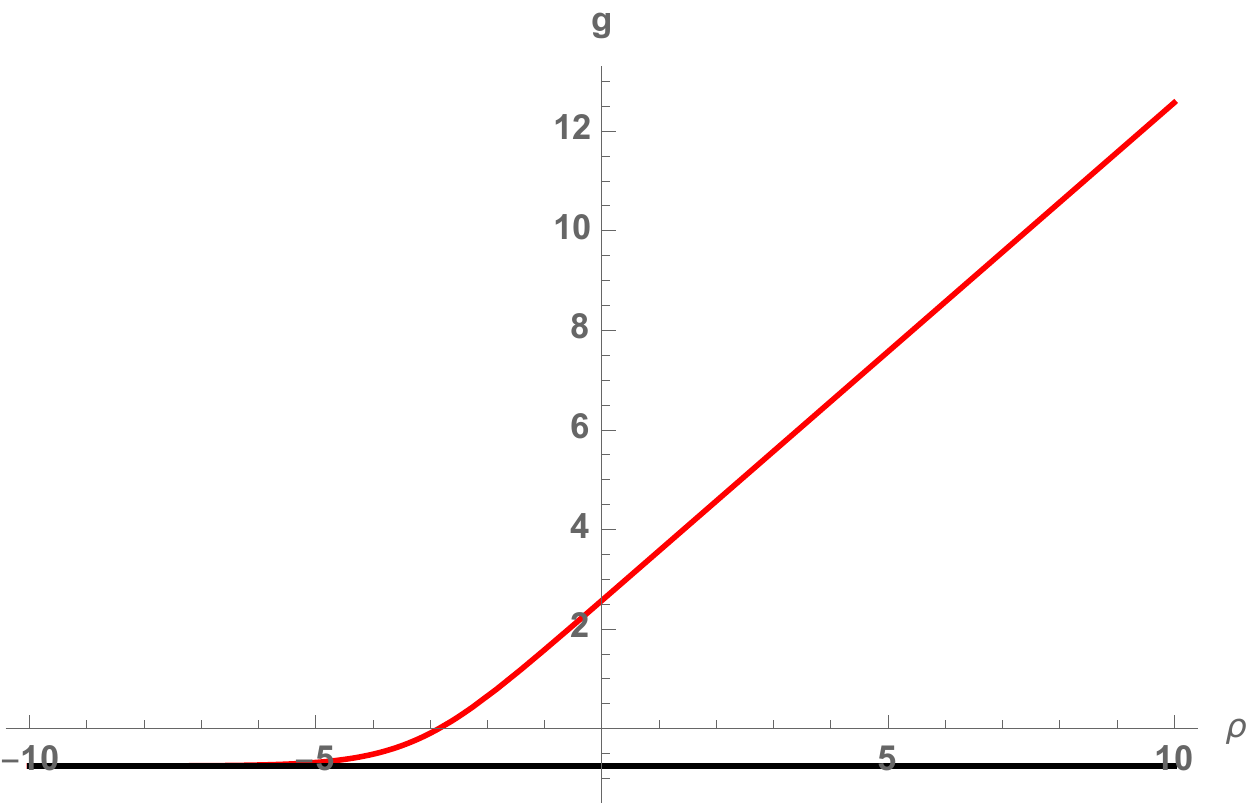} \qquad \includegraphics[width=2.0in]{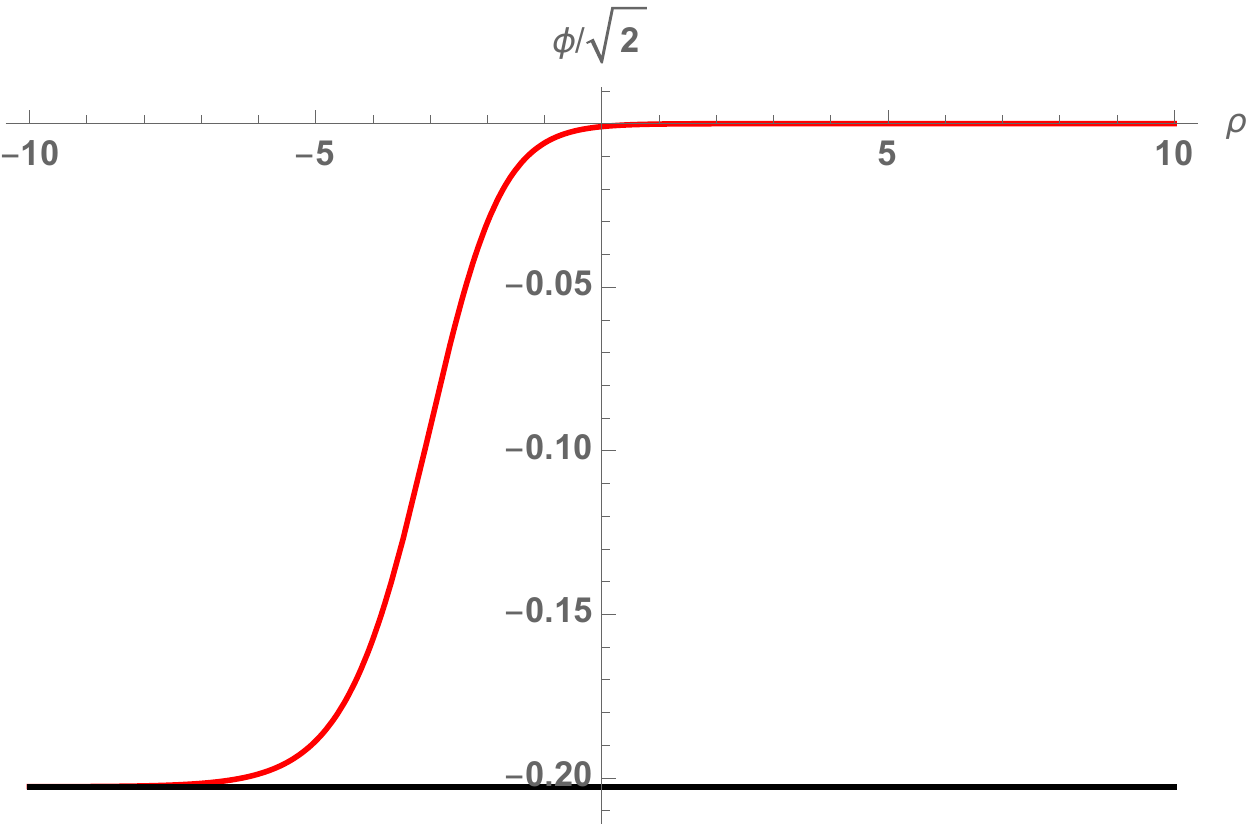}
\caption{{\it Numerical black hole solution with $m\,=\,\sqrt{2}$ and $g\,=\,3m$. The black straight lines are the values at the $AdS_2$ horizon given in \eqref{cayleyads2}.}}
\label{1}
\end{center}
\end{figure}

\bigskip

\section{Conclusions}

In this paper, we studied supersymmetric $AdS_6$ black holes with various horizon geometries. We found asymptotically $AdS_6$ black holes with $AdS_2\,\times\,\Sigma_{\mathfrak{g}_1}\times\Sigma_{\mathfrak{g}_2}$ horizon with $\mathfrak{g}_1>1$ and $\mathfrak{g}_2>1$ and presented the full black hole solution numerically. We calculated the Bekenstein-Hawking entropy of the black holes, and found that it nicely matches the topologically twisted index of 5d $USp(2N)$ gauge theory on $\Sigma_{\mathfrak{g}_1}\times\Sigma_{\mathfrak{g}_2}\times{S}^1$ in the large $N$ limit obtained in \cite{Hosseini:2018uzp}. The contribution to the topologically twisted index with zero magnetic fluxes, $\mathfrak{m}\,=\,0$, calulated in \cite{Crichigno:2018adf}, counts only the half of the gravitational entropy. We also found black hole solutions by considering D4-branes wrapped on K\"ahler four-cycles in Calabi-Yau fourfolds and Cayley four-cycles in $Spin(7)$ manifolds. 

For the near-horizon geometries we considered in this paper, we have shown that there should be non-zero two-form gauge potential, $B_{\mu\nu}$, in order to have a consistent set of supersymmetry equations, and satisfy the equations of motion. For this reason, as explained in the introduction, we concluded that the equations and solutions for the corresponding cases in the previous study, \cite{Naka:2002jz}, are $not$ correct.

It would be interesting to study supersymmetric four-cycles in matter coupled $F(4)$ gauged supergravity \cite{Andrianopoli:2001rs} along the line of \cite{Karndumri:2015eta}, to obtain supersymmetric black holes in $AdS_6$ \cite{Hosseini:2018usu, Suh:2018szn}.

\bigskip
\medskip
\leftline{\bf Acknowledgements}
\noindent We would like to thank Heeyeon Kim, Hyojoong Kim and Nakwoo Kim for very helpful discussions. We also would like to thank P. Marcos Crichigno, Alberto Zaffaroni and their collaborators of \cite{Hosseini:2018uzp} and \cite{Crichigno:2018adf} for helpful communications and kindly sharing the results in the revised versions of their manuscripts before publication. We are grateful to P. Marcos Crichigno, Nakwoo Kim, Hyojoong Kim, and Alberto Zaffaroni for invaluable comments on the preprint. This research was supported by the National Research Foundation of Korea under the grant NRF-2017R1D1A1B03034576.

\appendix
\section{The equations of motion}
\renewcommand{\theequation}{A.\arabic{equation}}
\setcounter{equation}{0} 

In this appendix, we present the equations of motion of $F(4)$ gauged supergravity,
\begin{align}
&R_{\mu\nu}\,=\,2\partial_\mu\phi\partial_\nu\phi+\frac{1}{8}g_{\mu\nu}\left(g^2e^{\sqrt{2}\phi}+4gme^{-\sqrt{2}\phi}-m^2e^{-3\sqrt{2}\phi}\right)-2e^{-\sqrt{2}\phi}\left(\mathcal{H}_\mu\,^\rho\mathcal{H}_{\nu\rho}-\frac{1}{8}g_{\mu\nu}\mathcal{H}_{\rho\sigma}\mathcal{H}^{\rho\sigma}\right) \notag \\ 
& \,\,\,\,\,\,\,\,\,\,\,\,\,\,\,\,\,\,\,\,\,\,\,\,\,\, -2e^{-\sqrt{2}\phi}\left(F^I_\mu\,^\rho{F}^I_{\nu\rho}-\frac{1}{8}g_{\mu\nu}F^I_{\rho\sigma}F^{I\rho\sigma}\right)+e^{2\sqrt{2}\phi}\left(G_\mu\,^{\rho\sigma}G_{\nu\rho_\sigma}-\frac{1}{6}g_{\mu\nu}G_{\rho\sigma\tau}G^{\rho\sigma\tau}\right)\,, \\
&\frac{1}{\sqrt{-g}}\partial_\mu\left(\sqrt{-g}g^{\mu\nu}\partial_\nu\phi\right)\,=\,\frac{1}{4\sqrt{2}}\left(g^2e^{\sqrt{2}\phi}-4gme^{-\sqrt{2}\phi}+3m^2e^{-3\sqrt{2}\phi}\right) \notag \\ 
& \,\,\,\,\,\,\,\,\,\,\,\,\,\,\,\,\,\,\,\,\,\,\,\,\,\, +\frac{1}{2\sqrt{2}}e^{-\sqrt{2}\phi}\left(\mathcal{H}_{\mu\nu}\mathcal{H}^{\mu\nu}+F^I_{\mu\nu}F^{I\mu\nu}\right)+\frac{1}{3\sqrt{2}}e^{2\sqrt{2}\phi}G_{\mu\nu\rho}G^{\mu\nu\rho}\,, \\
&\mathcal{D}_\nu\left(e^{-\sqrt{2}\phi}\mathcal{H}^{\nu\mu}\right)\,=\,\frac{1}{6}e\epsilon^{\mu\nu\rho\sigma\tau\kappa}\mathcal{H}_{\nu\rho}G_{\sigma\tau\kappa}\,, \\
&\mathcal{D}_\nu\left(e^{-\sqrt{2}\phi}F^{I\nu\mu}\right)\,=\,\frac{1}{6}e\epsilon^{\mu\nu\rho\sigma\tau\kappa}F^I_{\nu\rho}G_{\sigma\tau\kappa}\,, \\
&\mathcal{D}_\rho\left(e^{2\sqrt{2}\phi}G^{\rho\mu\nu}\right)\,=\,-\frac{1}{4}e\epsilon^{\mu\nu\rho\sigma\tau\kappa}\left(\mathcal{H}_{\rho\sigma}\mathcal{H}_{\tau\kappa}+F^I_{\rho\sigma}F^I_{\tau\kappa}\right)-me^{-\sqrt{2}\phi}\mathcal{H}^{\mu\nu}\,.
\end{align}




\begin{thebibliography}{99}

\bibitem{Maldacena:1997re}
  J.~M.~Maldacena,
  {\it The large N limit of superconformal field theories and supergravity,}
  Adv.\ Theor.\ Math.\ Phys.\  {\bf 2}, 231 (1998)
  [Int.\ J.\ Theor.\ Phys.\  {\bf 38}, 1113 (1999)]
  [arXiv:hep-th/9711200].

\bibitem{Romans:1985tw} 
  L.~J.~Romans,
  {\it The F(4) Gauged Supergravity in Six-dimensions,}
  Nucl.\ Phys.\ B {\bf 269}, 691 (1986).

\bibitem{Ferrara:1998gv} 
  S.~Ferrara, A.~Kehagias, H.~Partouche and A.~Zaffaroni,
  {\it AdS(6) interpretation of 5-D superconformal field theories,}  Phys.\ Lett.\ B {\bf 431}, 57 (1998)  [hep-th/9804006].

\bibitem{Seiberg:1996bd} 
  N.~Seiberg,
  {\it Five-dimensional SUSY field theories, nontrivial fixed points and string dynamics,}  Phys.\ Lett.\ B {\bf 388}, 753 (1996)  [hep-th/9608111].

\bibitem{Intriligator:1997pq} 
  K.~A.~Intriligator, D.~R.~Morrison and N.~Seiberg,
  {\it Five-dimensional supersymmetric gauge theories and degenerations of Calabi-Yau spaces,}  Nucl.\ Phys.\ B {\bf 497}, 56 (1997)  [hep-th/9702198].

\bibitem{Cvetic:1999un} 
  M.~Cvetic, H.~Lu and C.~N.~Pope,
  {\it Gauged six-dimensional supergravity from massive type IIA,}
  Phys.\ Rev.\ Lett.\  {\bf 83}, 5226 (1999) [hep-th/9906221].

\bibitem{Romans:1985tz} 
  L.~J.~Romans,
  {\it Massive N=2a Supergravity in Ten-Dimensions,}
  Phys.\ Lett.\ B {\bf 169}, 374 (1986)
  [Phys.\ Lett.\  {\bf 169B}, 374 (1986)].

\bibitem{Brandhuber:1999np} 
  A.~Brandhuber and Y.~Oz,
  {\it The D-4 - D-8 brane system and five-dimensional fixed points,}  Phys.\ Lett.\ B {\bf 460}, 307 (1999)  [hep-th/9905148].

\bibitem{Jeong:2013jfc} 
  J.~Jeong, O.~Kelekci and E.~O Colgain,
  {\it An alternative IIB embedding of F(4) gauged supergravity,}
  JHEP {\bf 1305}, 079 (2013) [arXiv:1302.2105 [hep-th]].

\bibitem{Hong:2018amk} 
  J.~Hong, J.~T.~Liu and D.~R.~Mayerson,
  {\it Gauged Six-Dimensional Supergravity from Warped IIB Reductions,}
  arXiv:1808.04301 [hep-th].

\bibitem{Malek:2018zcz} 
  E.~Malek, H.~Samtleben and V.~Vall Camell,
  {\it Supersymmetric AdS$_{7}$ and AdS$_6$ vacua and their minimal consistent truncations from exceptional field theory,}
  arXiv:1808.05597 [hep-th].

\bibitem{Maldacena:2000mw} 
  J.~M.~Maldacena and C.~Nunez,
  {\it Supergravity description of field theories on curved manifolds and a no go theorem,}
  Int.\ J.\ Mod.\ Phys.\ A {\bf 16}, 822 (2001) [hep-th/0007018].

\bibitem{Nunez:2001pt} 
  C.~Nunez, I.~Y.~Park, M.~Schvellinger and T.~A.~Tran,
  {\it Supergravity duals of gauge theories from F(4) gauged supergravity in six-dimensions,}
  JHEP {\bf 0104}, 025 (2001) [hep-th/0103080].

\bibitem{Naka:2002jz} 
  M.~Naka,
  {\it Various wrapped branes from gauged supergravities,}
  hep-th/0206141.

\bibitem{Dibitetto:2018iar} 
  G.~Dibitetto and N.~Petri,
  {\it Surface defects in the D4 $-$ D8 brane system,}
  arXiv:1807.07768 [hep-th].

\bibitem{Cacciatori:2009iz} 
  S.~L.~Cacciatori and D.~Klemm,
  {\it Supersymmetric AdS(4) black holes and attractors,}
  JHEP {\bf 1001}, 085 (2010) [arXiv:0911.4926 [hep-th]].

\bibitem{Benini:2015noa} 
  F.~Benini and A.~Zaffaroni,
  {\it A topologically twisted index for three-dimensional supersymmetric theories,}
  JHEP {\bf 1507}, 127 (2015) [arXiv:1504.03698 [hep-th]].

\bibitem{Benini:2015eyy} 
  F.~Benini, K.~Hristov and A.~Zaffaroni,
  {\it Black hole microstates in AdS$_{4}$ from supersymmetric localization,}
  JHEP {\bf 1605}, 054 (2016) [arXiv:1511.04085 [hep-th]].

\bibitem{Hosseini:2018uzp} 
  S.~M.~Hosseini, I.~Yaakov and A.~Zaffaroni,
  {\it Topologically twisted indices in five dimensions and holography,}
  JHEP {\bf 1811}, 119 (2018) [arXiv:1808.06626 [hep-th]].

\bibitem{Crichigno:2018adf} 
  P.~M.~Crichigno, D.~Jain and B.~Willett,
  {\it 5d Partition Functions with A Twist,}
  JHEP {\bf 1811}, 058 (2018) [arXiv:1808.06744 [hep-th]].

\bibitem{Gauntlett:2001jj} 
  J.~P.~Gauntlett and N.~Kim,
  {\it M five-branes wrapped on supersymmetric cycles. 2.,}
  Phys.\ Rev.\ D {\bf 65}, 086003 (2002) [hep-th/0109039].

\bibitem{Benini:2013cda} 
  F.~Benini and N.~Bobev,
  {\it Two-dimensional SCFTs from wrapped branes and c-extremization,}
  JHEP {\bf 1306}, 005 (2013) [arXiv:1302.4451 [hep-th]].

\bibitem{Gauntlett:2000ng} 
  J.~P.~Gauntlett, N.~Kim and D.~Waldram,
  {\it M Five-branes wrapped on supersymmetric cycles,}
  Phys.\ Rev.\ D {\bf 63}, 126001 (2001) [hep-th/0012195].

\bibitem{Bobev:2017uzs} 
  N.~Bobev and P.~M.~Crichigno,
  {\it Universal RG Flows Across Dimensions and Holography,}
  JHEP {\bf 1712}, 065 (2017) [arXiv:1708.05052 [hep-th]].

\bibitem{Jafferis:2012iv} 
  D.~L.~Jafferis and S.~S.~Pufu,
  {\it Exact results for five-dimensional superconformal field theories with gravity duals,}
  JHEP {\bf 1405}, 032 (2014) [arXiv:1207.4359 [hep-th]].

\bibitem{Gauntlett:1998vk} 
  J.~P.~Gauntlett, N.~D.~Lambert and P.~C.~West,
  {\it Branes and calibrated geometries,}
  Commun.\ Math.\ Phys.\  {\bf 202}, 571 (1999) [hep-th/9803216].

\bibitem{Andrianopoli:2001rs} 
  L.~Andrianopoli, R.~D'Auria and S.~Vaula,
  {\it Matter coupled F(4) gauged supergravity Lagrangian,}
  JHEP {\bf 0105}, 065 (2001) [hep-th/0104155].

\bibitem{Karndumri:2015eta} 
  P.~Karndumri,
  {\it Twisted compactification of N = 2 5D SCFTs to three and two dimensions from F(4) gauged supergravity,}
  JHEP {\bf 1509}, 034 (2015) [arXiv:1507.01515 [hep-th]].

\bibitem{Hosseini:2018usu} 
  S.~M.~Hosseini, K.~Hristov, A.~Passias and A.~Zaffaroni,
  {\it 6D attractors and black hole microstates,}
  arXiv:1809.10685 [hep-th].

\bibitem{Suh:2018szn} 
  M.~Suh,
  {\it Supersymmetric $AdS_6$ black holes from matter coupled $F(4)$ gauged supergravity,}
  arXiv:1810.00675 [hep-th].

\end{thebibliography}
\end{document}